\documentclass[11pt,a4paper]{article}
\usepackage[margin=1in]{geometry}
\usepackage{graphicx,comment,sansmath,amssymb,amsmath,setspace,authblk,cite}
\usepackage[usenames,dvipsnames,svgnames,table]{xcolor}
\usepackage[font={small,sf},labelfont={bf}]{caption}
\usepackage[font={small,sf},labelfont={bf}]{subcaption}
\usepackage{pdfpages}
\DeclareCaptionFont{sansmath}{\sansmath}
\newcommand{\figautomata}{1}
\newcommand{\figconvergence}{3}
\newcommand{\figerrorrates}{4}
\newcommand{\SI}{Supplementary Information}

\title{Forgiver triumphs in alternating Prisoner's Dilemma}
\author[1]{Benjamin M. Zagorsky\thanks{These authors contributed equally to this work.}}
\author[2]{Johannes G. Reiter$^{*}$}
\author[2]{Krishnendu Chatterjee}
\author[1,3]{Martin A. Nowak\thanks{\footnotesize{To whom correspondence should be addressed. E-mail: martin\_nowak@harvard.edu (M.A.N.)}}}
\affil[1]{Program for Evolutionary Dynamics, Harvard University, Cambridge, Massachusetts 02138, USA}
\affil[2]{IST Austria (Institute of Science and Technology Austria), Klosterneuburg 3400, Austria}
\affil[3]{Department of Mathematics and Department of Organismic and Evolutionary Biology, Harvard University, Cambridge, Massachusetts 02138, USA}
\date{}

\onehalfspace

\begin{document}
\maketitle

\begin{abstract}
Cooperative behavior, where one individual incurs a cost to help another, is a wide spread  phenomenon. Here we study direct reciprocity in the context of the alternating Prisoner's Dilemma. We consider all strategies that can be implemented by one and two-state automata. We calculate the payoff matrix of all pairwise encounters in the presence of noise. We explore deterministic selection dynamics with and without mutation. Using different error rates and payoff values, we observe convergence to a small number of distinct equilibria. Two of them are uncooperative strict Nash equilibria representing always-defect (ALLD) and Grim. The third equilibrium is mixed and represents a cooperative alliance of several strategies, dominated by a strategy which we call Forgiver. Forgiver cooperates whenever the opponent has cooperated; it defects once when the opponent has defected, but subsequently Forgiver attempts to re-establish cooperation even if the opponent has defected again. Forgiver is not an evolutionarily stable strategy, but the alliance, which it rules, is asymptotically stable. For a wide range of parameter values the most commonly observed outcome is convergence to the mixed equilibrium, dominated by Forgiver. Our results show that although forgiving might incur a short-term loss it can lead to a long-term gain. Forgiveness facilitates stable cooperation in the presence of exploitation and noise.
\end{abstract}

A cooperative dilemma arises when two cooperators receive a higher payoff than two defectors and yet there is an incentive to defect\cite{Hauert2006synergy,Nowak2012evolving}. The Prisoner's Dilemma\cite{Rapoport1965,Trivers1971,Axelrod1984,May1987more,Milinski1987tit,Sigmund1993games,Clutton2009cooperation} is the strongest form of a cooperative dilemma, where cooperation requires a mechanism for its evolution\cite{Nowak2006five}.
Direct reciprocity is a mechanism for the evolution of cooperation based on repeated interactions between the same two individuals\cite{Aumann1981survey,Fudenberg1986,Abreu1988theory,Kraines1989pavlov,Nowak1989oscillations,Fudenberg1990,Nowak1990stochastic,Nowak1992tit,Nowak1993strategy,Wu1995cope,Kraines1995evolution,Cressman1996evolutionary,Boerlijst1997,Milinski1998working,Kraines2000natural,Nowak2004emergence,DalBo2005cooperation,Doebeli2005models,Kendall2007iterated,Sigmund2009calculus,Imhof2010stochastic,Fudenberg2012slow,VanVeelen2012direct,Press2012iterated,Hilbe2013evolution,Stewart2013extortion}. The standard theory assumes that both players decide simultaneously what do for the next round. But another possibility is that the players take turns when making their moves\cite{Wedekind1996human,Nowak1994alternating,Frean1994prisoner}. 
This implementation can lead to a strictly alternating game, where the players always chose their moves in turns, or a stochastically alternating game, where in each round the player to move next is selected probabilistically. Here we investigate the strictly alternating game.

We consider the following scenario. In each round a player can pay a cost, $c$, for the other player to receive a benefit, $b$, where $b>c>0$. If both players cooperate in two consecutive moves, each one gets $b-c$, which is greater than the zero payoff they would receive for mutual defection. But if one player defects, while the other cooperates, then the defector gets payoff, $b$, while the cooperator gets the lowest payoff, $-c$. Therefore, over two consecutive moves the payoff structure is the same as in a Prisoner's Dilemma: $b>b-c>0>-c$. Thus, this game is an alternating Prisoner's Dilemma\cite{Nowak1994alternating}. 

We study the strictly alternating Prisoner's Dilemma in the presence of noise\cite{Fudenberg1990,Nowak1995automata,Wu1995cope}.
In each round, a player makes a mistake with probability $\epsilon$ leading to the opposite move. We consider all strategies that can be implemented by deterministic finite state automata\cite{Hopcroft2006} with one or two states. Finite state automata have been used extensively to study repeated games\cite{Rubinstein1986finite,Miller1996coevolution,Binmore1992evolutionary,Nowak1995automata} including the simultaneous Prisoner's Dilemma. 
In our case, each state is labeled by $C$ or $D$. In state $C$ the player will cooperate in the next move; in state $D$ the player will defect. Each strategy starts in one of those two states.
Each state has two outgoing transitions (either to the same or to the other state): one transition specifies what happens if the opponent has cooperated (labeled with $c$) and one if the opponent has defected (labeled with $d$). There are 26 automata encoding unique strategies (Fig. \figautomata). These strategies include ALLC, ALLD,  Grim, tit-for-tat (TFT), and win-stay lose-shift (WSLS).

We can divide these 26 strategies into four categories: (i) sink-state C (ssC) strategies, (ii) sink-state D (ssD) strategies, (iii) suspicious dynamic strategies, and (iv) hopeful dynamic strategies. Sink-state strategies always-cooperate or always-defect either from the beginning or after some condition is met. They include ALLC, ALLD, Grim and variations of them. There are eight sink-state strategies in total. Suspicious dynamic strategies start with defection and then move between their defective and cooperative state depending on the other player's decision. Hopeful dynamic strategies do the same, but start with cooperation. There are nine strategies in each of these two categories. For each suspicious dynamic strategy there is a hopeful counterpart.

Some of the dynamic strategies do little to optimize their score. 
For example, Alternator ($S_{22}$; see Fig. \figautomata\ for strategy names and their indexing) switches between cooperation and defection on each move. But a subset of dynamic strategies are of particular interest: Forgiver ($S_{14}$), TFT ($S_{15}$), WSLS ($S_{16}$) and their suspicious counterparts ($S_{4}$, $S_{8}$, and $S_{12}$). These strategies have the design element to stay in state $C$ if the opponent has cooperated in the last round but move to state $D$ if the opponent has defected:
\begin{equation}
\includegraphics[scale=1.3]{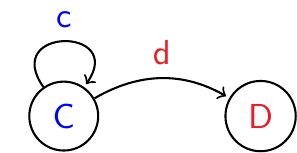}
\label{figelement}
\end{equation}
We call this element (\ref{figelement}) the conditional cooperation element.
TFT then requires the opponent to cooperate again in order to move back to the cooperative state. WSLS in contrast requires the opponent to defect in order to move back to the cooperative state. But Forgiver moves back to the cooperative state irrespective of the opponents move (Fig.~1: hopeful dynamic strategies). 

Neither TFT nor WSLS are error correcting in the alternating game\cite{Nowak1994alternating}. In a game between two TFT players, if by mistake one of them starts to defect, they will continue to defect until another mistake happens. The same is true for WSLS in the alternating game. In the simultaneous game TFT is also not error correcting\cite{Nowak1992tit}, while WSLS is error correcting\cite{Nowak1993strategy}. Thus WSLS, which is known to be a strong strategy in the simultaneous game, is not expected to do well in the alternating game. Forgiver, on the other hand, is error correcting in the alternating game. It recovers from an accidental defection in three rounds (Fig. 2).

\section*{Results}
We calculate the payoff for all pairwise encounters in games of $L$ moves of both strategies, thereby obtaining a $26\times 26$ payoff matrix. 
We average over which strategy goes first. 
Without loss of generality we set $c=1$.
At first we study the case $b=3$ with error rate $\epsilon=0.05$ and game length $L=100$.
Table \ref{tbl:payoff3_L=100} shows a part of the calculated payoff matrix for six relevant strategies. 
We find that ALLD ($S_{1}$) and Grim ($S_{17}$) are the only strict Nash equilibria among the 26 pure strategies. 
ALLC ($S_{26}$) vs ALLC receives a high payoff, but so does Forgiver vs Forgiver. The payoffs of WSLS vs WSLS and TFT vs TFT are low, because neither strategy is error correcting (Fig. 2). Interestingly TFT vs WSLS yields good payoff for both strategies, because their interaction is error correcting.
\begin{table}[b]
\centering
\begin{tabular}{ c | c c c c c c }
	 & \textbf{ALLD}	& \textbf{Forgiver}	& \textbf{TFT}	& \textbf{WSLS}	& \textbf{Grim}	& \textbf{ALLC}  \\ \hline
\textbf{ALLD}  & \textcolor{red}{10.0} & 148.4 & 24.8 & 144.9 & 11.5 & 280.0 \\ 
\textbf{Forgiver}  & -36.1 & 174.8 & 163.5 & 166.9 & -12.7 & 194.3 \\ 
\textbf{TFT}  & 5.1 & 178.1 & 104.5 & 176.7 & 24.5 & 194.5 \\ 
\textbf{WSLS}  & -35.0 & 169.1 & 162.3 & 106.5 & -12.0 & 230.9 \\ 
\textbf{Grim}  & 9.5 & 152.0 & 40.5 & 148.8 & \textcolor{red}{28.1} & 262.9 \\ 
\textbf{ALLC}  & -80.0 & 177.2 & 176.6 & 67.3 & -28.8 & 190.0 \\ 
	 
  \end{tabular}
\caption{Excerpt of the payoff matrix with the most relevant strategies when the benefit value $b=3$ ($c=1$), the error rate $\epsilon = 5$\%, and the number of rounds in each game $L=100$. 
There are two pure Nash equilibria in the full payoff matrix: ALLD ($S_{1}$) and Grim ($S_{17}$), both denoted in red.}
\label{tbl:payoff3_L=100}
\end{table}

In the following, we study evolutionary game dynamics\cite{Hofbauer1988, Hofbauer1998,Weibull1995} with the replicator equation. The frequency of strategy $S_i$ is denoted by $x_i$.  At any one time we have $\sum_{i=1}^n x_i=1$, where $n=26$ is the number of strategies. The frequency $x_i$ changes according to the relative payoff of strategy~$S_i$.  We evaluate evolutionary trajectories for many different initial frequencies. The trajectories start from $10^4$ uniformly distributed random points in the $26$-simplex. 

Typically, we do not find convergence to one of the strict Nash equilibria (Fig.~\figconvergence b). 
In only 5\% of the cases the trajectories converge to the pure ALLD equilibrium and in 18\% of the cases the trajectories converge to the pure Grim equilibrium.
However, in 77\% of the cases we observe convergence to a mixed equilibrium of several strategies, dominated by Forgiver with a population share of 82.6\% (Fig.~\figconvergence b). 
The other six strategies present in this cooperative alliance are Paradoxic Grateful ($S_{5}$; population share of~3.2\%), Grateful ($S_{9}$;~5.6\%), Suspicious ALLC ($S_{13}$;~3.8\%), and ALLC ($S_{26}$;~0.3\%), all of which have a sink-state C,  and TFT ($S_{15}$; 4.1\%) and WSLS ($S_{16}$; 0.4\%), which are the remaining two dynamic strategies with the conditional cooperation element. 

When increasing the benefit value to $b=4$ and $b=5$, we observe convergence to a very similar alliance (Fig.~\figconvergence c and \figconvergence d). 
For $b = 2$, however, the ssC (sink-state C) strategies ($S_{5}$, $S_{9}$, $S_{13}$, $S_{26}$) and WSLS are replaced by Grim and the mixed equilibrium is formed by Forgiver, TFT, and Grim (Fig.~\figconvergence a). 
Very rarely we observe convergence to a cooperative alliance led by Suspicious Forgiver ($S_{12}$; for short, sForgiver).
It turns out that for some parameter values the Suspicious Forgiver alliance is an equilibrium (see \SI).

From the $10^4$ random initial frequencies, the four equilibria were reached in the following proportions (using $\epsilon=0.05$ and $L=100$):
\begin{center}
  \begin{tabular}{ c | c c c c}
     & \textbf{ALLD}  & \textbf{Grim} & \textbf{Forgiver} & \textbf{sForgiver} \\ \hline
     \textbf{b = 2} & 15\% & 52\% &	33\% & $<$1\% \\ 
     \textbf{b = 3} & 5\% &	18\% & 77\% & 0\% \\ 
		 \textbf{b = 4} & 2\% & 7\% &	90\% & 1\% \\ 
     \textbf{b = 5} & 1\% &	3\% &	93\% & 3\%.\\ 
  \end{tabular}
\end{center}

Changing the error rate, $\epsilon=0.01, 0.05, 0.1$ and the number of rounds per game, $L=10,$ $100, 1000$, we find very similar behavior.
Only the frequencies of the strategies within the mixed equilibria change marginally but not the general equilibrium composition (Fig. \figerrorrates). 
Though, there is one exception.
When the probability for multiple errors within an entire match becomes very low (e.g., $L=10$ and $\epsilon = 0.05$ or $L=100$ and $\epsilon=0.01$) and $b>2$, the payoff of ALLC against Grim can become higher than the payoff of Grim against itself. In other words, Grim can be invaded by ALLC. Hence, instead of the pure Grim equilibrium we observe a mixed equilibrium between Grim and ALLC (\SI).

We check the robustness of the observed equilibria by incorporating mutation to the replicator equation. We find that both the ALLD and the rare Suspicious Forgiver equilibrium are unstable. In the presence of mutation the evolutionary trajectories lead away from ALLD to Grim and from Suspicious Forgiver to Forgiver (see \SI).
The Grim equilibrium and the Forgiver equilibrium remain stable. 

Essential is that Forgiver can resist invasion by ssD strategies ($S_{1}$, $S_{17}$, $S_{21}$, $S_{25}$), because Forgiver does better against itself than the ssD strategies do against Forgiver (Table \ref{tbl:payoff3_L=100}).
However, Forgiver can be invaded by ssC strategies and TFT. But, since TFT performs poorly against itself and ssC strategies are exploited by WSLS (Table \ref{tbl:payoff3_L=100}), all these strategies can coexist in the Forgiver equilibrium.
More detailed results and equilibrium analysis for a wide range of parameter values for $\epsilon$ and $L$ are provided in the \SI .

In the case of an infinitely repeated game we can derive analytical results for the average payoff per round for the most relevant strategy pairs (Table \ref{tbl:analyticalresults}).
From these results we obtain that ALLD (or ssD strategies) cannot invade Forgiver if
\begin{align}
\frac{b}{c} > \frac{2+\epsilon-\epsilon^2}{1-2\epsilon} \ .
\label{eq:invasion}
\end{align}
This result holds for any error rate, $\epsilon$, between $0$ and $1/2$ (Fig. \figerrorrates d).
\begin{table}[b]%
\centering
\begin{tabular}{ c | c c c }
 	& \textbf{ALLD}	& \textbf{Forgiver}	& \textbf{ALLC}  \\ \hline
\textbf{ALLD} & $\epsilon\cdot(b-c)$ & $b \cdot\frac{1-\epsilon^2}{2-\epsilon}-\epsilon \cdot c$ & $b - \epsilon \cdot (b+c)$\\
\textbf{Forgiver}	& $\epsilon  \cdot b-c \cdot \frac{1-\epsilon^2}{2-\epsilon}$ & $(b-c)  \cdot \frac{1+\epsilon^2 \cdot (1-\epsilon)}{1+3\epsilon-\epsilon^2}$ & $b \cdot (1-\epsilon)-c  \cdot \frac{1-\epsilon \cdot(1-\epsilon)}{1+\epsilon}$ \\
\textbf{ALLC} & $\epsilon  \cdot b - c \cdot(1-\epsilon)$ & $b  \cdot \frac{1-\epsilon+\epsilon^2}{1+\epsilon}-c \cdot(1-\epsilon) $ & $(b-c)(1-\epsilon)$ \\
\end{tabular}
\caption{Analytical results of the average payoff per round in the infinitely alternating Prisoner's Dilemma for ALLD ($S_{1}$), Forgiver ($S_{14}$), and ALLC ($S_{26}$) playing against each other. Derivations are provided in the \SI .}
\label{tbl:analyticalresults}
\end{table}

\section*{Discussion}
Our results imply an indisputable strength of the strategy Forgiver in the alternating Prisoner's Dilemma in the presence of noise.
For a wide range of parameter values, Forgiver is the dominating strategy of the cooperative equilibrium, having a population share of more than half in all investigated scenarios.

Essential for the success of a cooperative strategy in the presence of noise is how fast it can recover back to cooperation after a mistake, but at the same time, also avoid excessive exploitation by defectors.
The conditional cooperation element (\ref{figelement}) is crucial for the triumph of Forgiver.
Even though, also TFT and WSLS contain this element, which allows them to cooperate against cooperative strategies without getting excessively exploited by defectors, these strategies are not  as successful as Forgiver, because of their inability to correct errors.
Grim also possesses this conditional cooperation element. 
However, noise on the part of Grim's opponent will inevitably cause Grim to switch to always-defect. It is Grim's ability to conditionally cooperate for the first handful of turns that provides a competitive advantage over pure ALLD such that the strict Nash equilibrium ALLD can only rarely arise.

The other strategies appearing in the Forgiver equilibrium for the cases of $b=3$, $b=4$, and $b=5$ are Paradoxic Grateful ($S_{5}$), Grateful ($S_{9}$), Suspicious ALLC ($S_{13}$), and ALLC ($S_{26}$). 
All of them are ssC strategies that, in the presence of noise, behave like ALLC after the first few moves.  The strategy ALLC does very well in combination with Forgiver. 
Nevertheless, ALLC itself appears rarely. Perhaps because of Paradoxic Grateful, which defects against ALLC for many moves in the beginning, whereas Suspicious ALLC puts Paradoxic Grateful into its cooperating state immediately. One might ask why these ssC strategies do not occupy a larger population share in the cooperative equilibrium. The reason is the presence of exploitative strategies like WSLS which itself is a weak strategy in this domain. 
If only Forgiver was present, WSLS would be quickly driven to extinction; WSLS does worse against itself and Forgiver than Forgiver does against WSLS and itself (see Table \ref{tbl:payoff3_L=100}). 
But WSLS remains in the Forgiver equilibrium because it exploits the ssC strategies.
Interestingly, higher error rates increase the population share of unconditional cooperators (ssC strategies) in the cooperative equilibrium (Fig. 4c). 
Simultaneously, the higher error rates can decrease the probability to converge to the cooperative equilibrium dramatically and hence prevent the evolution of any cooperative behavior (Fig. 4a).

Grim and Forgiver are similar strategies, the difference being, in the face of a defection, Forgiver quickly returns to cooperation whereas Grim never returns.  An interesting interpretation of the relationship is that Grim never forgives while Forgiver always does. Thus, the clash between Grim and Forgiver is actually a test of the viability of forgiveness under various conditions.  On the one hand, the presence of noise makes forgiveness powerful and essential.  On the other hand, if cooperation is not valuable enough, forgiveness can be exploited. Moreover, even when cooperation is valuable, but the population is ruled by exploiters, forgiveness is not a successful strategy. Given the right conditions, forgiveness makes cooperation possible in the face of both exploitation and noise.

These results demonstrate a game-theoretic foundation for forgiveness as a means of promoting cooperation. If cooperation is valuable enough, it can be worth forgiving others for past wrongs in order to gain future benefit. Forgiving incurs a short-term loss but ensures a greater long-term gain. Given all the (intentional or unintentional) misbehavior in the real world, forgiveness is essential for maintaining healthy, cooperative relationships.

\section*{Methods}

\textbf{Strategy space.}
We consider deterministic finite automata\cite{Hopcroft2006} (DFA) with one and two states.
There are two one-state automata which encode the strategies always-defect (ALLD) and always-cooperate (ALLC).
In total, there are 32 two-state automata encoding strategies in our game: two possible arrangements of states ($CD$, $DC$) and 16 possible arrangements of transitions per arrangement of states.
For 8 of these 32 automata, the second state is not reachable, making them indistinguishable from a one-state automata.
Since we already added the one-state automata to our strategy space, these 8 can be ignored.
The remaining 24 two-state automata encode distinct strategies in our game.
Hence, in total we have 26 deterministic strategies in the alternating Prisoner's Dilemma (Fig. 1).

\textbf{Generation of the payoff matrix.}
In each round of the game a player can either cooperate or defect. 
Cooperation means paying a cost, $c$, for the other player to receive a benefit, $b$. 
Defection means paying no cost and distributing no benefit. 
Thus, summing over two consecutive moves (equivalent to one round) we obtain the following payoff matrix:
\begin{center}
  \begin{tabular}{ c | c c }
     & \textbf{C} & \textbf{D} \\ \hline
    \textbf{C} & $b-c$, $b-c$ & $-c$, $b$ \\
    \textbf{D} & $b$, $-c$ & $0$, $0$ \ . \\
  \end{tabular}
\end{center}
If $b>c>0$, the game is a Prisoner's Dilemma since the following inequalities are satisfied: $b > b-c > 0 > -c$ and $2(b-c) > b-c$.
In other words, in a single round it is best to defect, but cooperation might be fruitful when playing over multiple rounds.
The second inequality ensures that sustained cooperation results in a higher payoff than alternation between cooperation and defection.

For each set of parameters (number of rounds $L$, error rate $\epsilon$, benefit value $b$, and costs $c$), we generate a $26\times 26$ payoff matrix $A$ where each of the 26 distinct strategies is paired with each other.
The entry $a_{ij}$ in the payoff matrix $A$ gives the payoff of strategy $S_i$ playing against strategy $S_j$.
Based on the average of which strategy (player) goes first, we define the initial state distribution of both players as a row vector $Q_{S_i \times S_j}$.
Since the players do not observe when they have made a mistake (i.e., the faulty player does not move to the corresponding state of the erroneous action which he has accidentally played), the state space consists of sixteen states namely $CC$, $CD$, $DC$, $DD$, $D^*C$, $D^*D$, $C^*C$, $C^*D$, $CD^*$, $\cdots$ $C^*C^*$. 
The star after a state indicates that the player accidentally played the opposite move as intended by her current state.

Each game consists of exactly $L$ moves of both player. 
In each move, a player makes a mistake with probability $\epsilon$ and thus implements the opposite move of what is specified by her strategy (automaton).	
We denote $1 - \epsilon$ by $\bar{\epsilon}$.
Although, the players do not observe their mistakes, the payoffs depend on the actual moves. The payoffs corresponding to their moves in the different states are given by the column vector $U$.
 
Next, we define a $16 \times 16$ transition matrix $M_{S_i \times S_j}$ for each pair of strategies $S_i$, $S_j$.
The entries of the transition matrix are given by the probabilities to move from one state to the next:
\begin{align}
M_{S_i \times S_j} &=
\left( 
\begin{array}{ccccc}
p_1 p_1'\bar{\epsilon}^2 & p_1(1-p_1')\bar{\epsilon}^2 & (1-p_1)p_1'\bar{\epsilon}^2 & \cdots &
(1-p_1)(1-p_1')\epsilon^2  \\
p_2 p_3'\bar{\epsilon}^2 & \ddots & & & \vdots \\
p_3 p_2'\bar{\epsilon}^2 & & & & \\
\vdots & & & \ddots & \vdots \\
p_3 p_3'\bar{\epsilon}^2 & \cdots & & \cdots & (1-p_3)(1-p_3') \epsilon^2 \\
\end{array} 
\right) 
\label{eq:transitionmatrix-noise}
\end{align}
where the quadruple\cite{Nowak1994alternating} $(p_1,p_2,p_3,p_4)$ defines the probabilities of strategy $S_i$ to cooperate in the observed states $CC$, $CD$, $DC$, and $DD$ (errors remain undetected by the players). 
Respectively, the quadruple $(p_1',p_2',p_3',p_4')$ encodes the strategy $S_j$.
For example, $(1-p_4)\epsilon p_3\bar{\epsilon}$ is the probability to move from state $DD$ to state $C^*C$. 
A deterministic strategy is represented as a quadruple where each ${p_l \in \{0,1\}}$. 

Using the initial state distribution $Q_{S_i \times S_j}$ , the transition matrix $M_{S_i \times S_j}$, and the payoff vector $U$, we calculate the payoff $a_{ij}$ of strategy $S_i$ playing against strategy $S_j$ via a Markov Chain:  
\begin{equation}
a_{ij}
= {Q}_{S_i \times S_j} \times \sum_{k=0}^{L-1}{{M}_{S_i \times S_j}^k} \  \times {U} \ .
\label{eq:reward}
\end{equation}
Applying equation (\ref{eq:reward}) to each pair of strategies, we obtain the entire payoff matrix $A$ for a given set of parameter values.
Although we use deterministic strategies, the presence of noise implies that the game that unfolds between any two strategies is described by a stochastic process.
Payoff matrices for benefit values of $b=2$, $b=3$, $b=4$, and $b=5$, for error rates of $\epsilon=0.01$, $\epsilon=0.05$, and $\epsilon=0.1$, and for game length of $L=10$, $L=100$, and $L=1000$ are provided in the \SI .

\textbf{Evolution of strategies.}
The strategy space spans a $26$-simplex which we explore via the replicator equation\cite{Hofbauer1988, Hofbauer1998, Weibull1995} with and without mutations.
The frequency of strategy $S_i$ is given by $x_i$. 
At any time $\sum_{i=1}^{n} x_i=1$ holds where $n=26$ is the number of strategies. 
The average payoff (fitness) for strategy $S_i$ is given by 
\begin{equation}
f_i=\sum_{j=1}^n a_{ij} x_j \ .
\label{eq:fitness}
\end{equation}
The frequency of strategy $S_i$ changes according to the differential equation 
\begin{equation}
\dot{x_i} = x_i (f_i - \bar{f}) + u \left(\frac{1}{n}-x_i \right)
\label{eq:replicator}
\end{equation}
where the average population payoff is $\bar f=\sum_{i=1}^n f_ix_i$ and $u$ is the mutation rate.
Using the differential equation (\ref{eq:replicator}), defined on the $n$-simplex (here, $n=26$), we study the evolutionary dynamics in the alternating Prisoner's dilemma for many different initial conditions (i.e., random initial frequencies of the strategies).
We generate a uniform-random point in the $n$-simplex by taking the negative logarithm of $n$ random numbers in $(0,1)$, then normalizing these numbers such that they sum to $1$, and using the normalized values as the initial frequencies of the $n$ strategies\cite{Devroye1986non}. 

\textbf{Computer simulations.}
Our computer simulations are implemented in Python and split into three programs.
The first program generates the $26\times 26$ payoff matrix for each set of parameters.
The second program simulates the deterministic selection dynamics starting from uniform-random points in the $26$-simplex.
The third program performs statistical analysis on the results of the second program.
The code is available upon request.

\bibliographystyle{naturemag}
\bibliography{egt}

\includepdf[pages={1}]{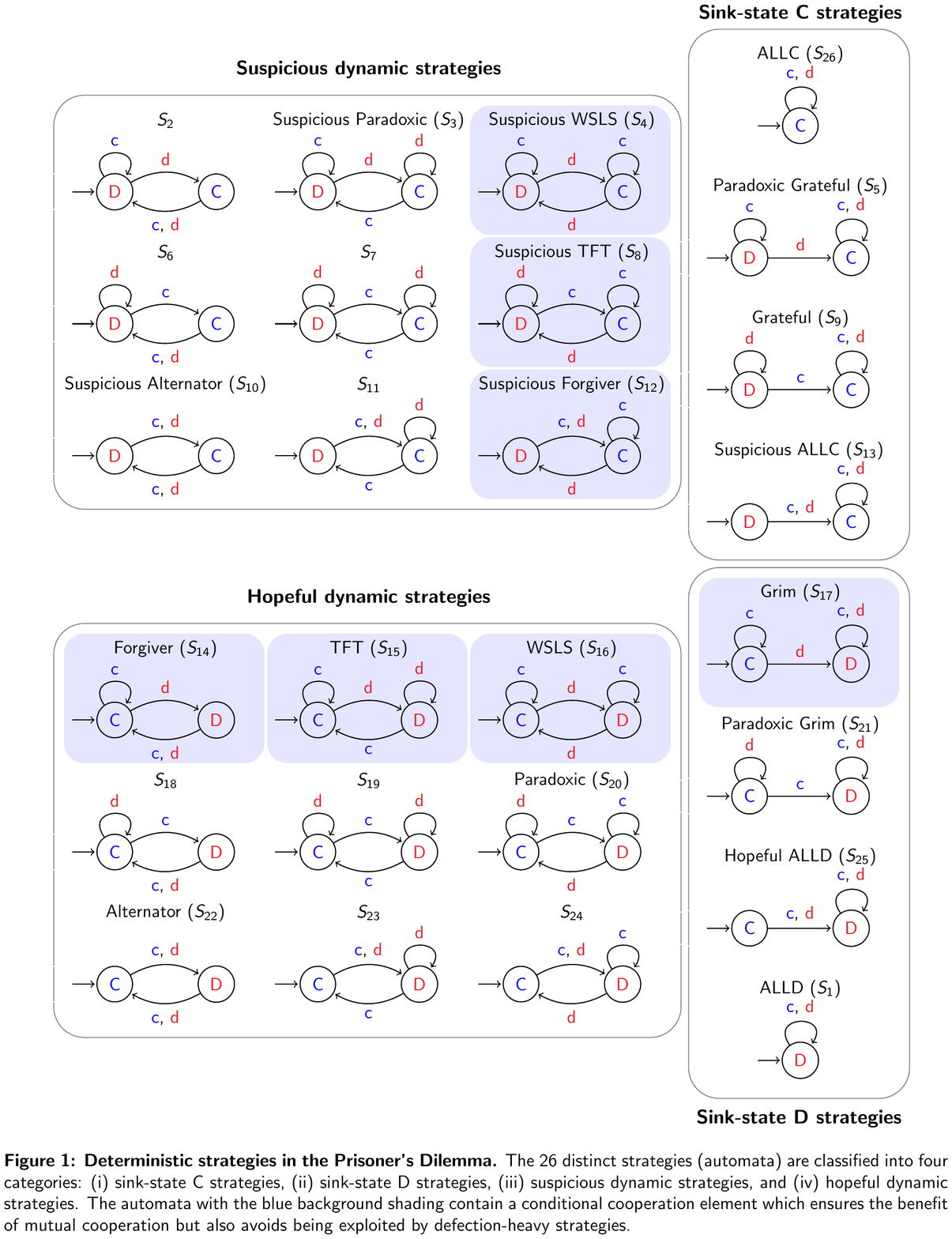}
\includepdf[pages={1}]{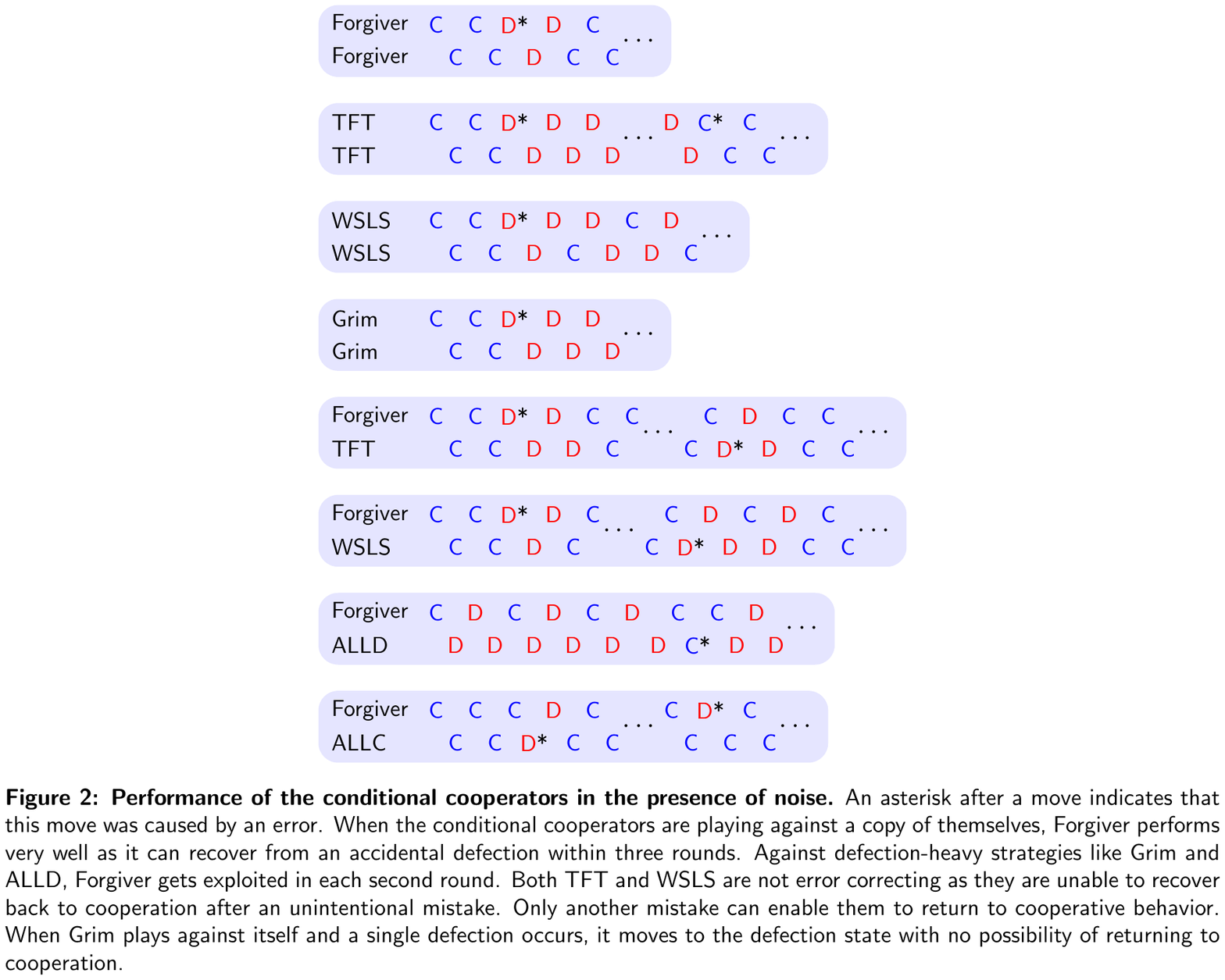}
\includepdf[pages={1}]{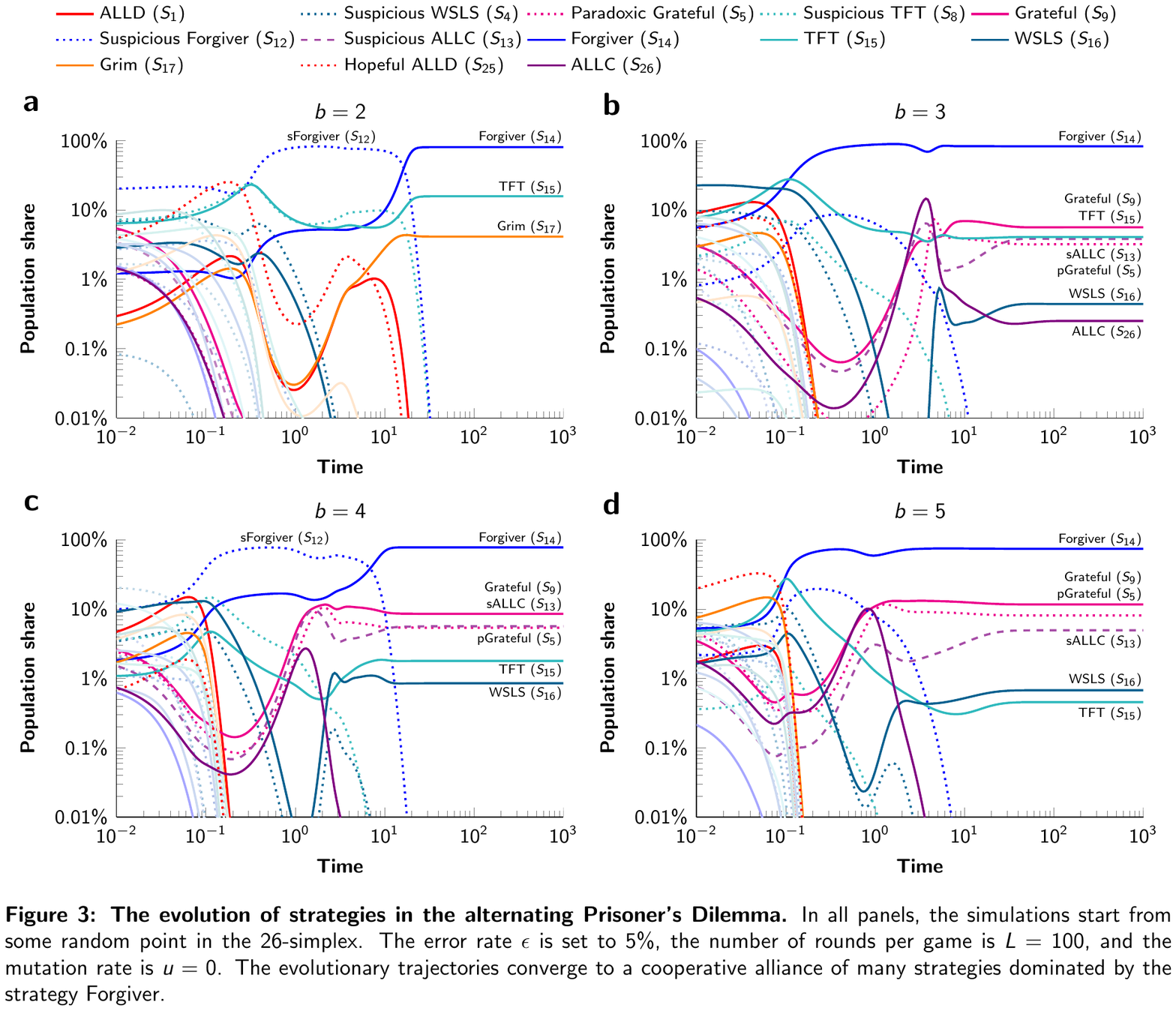}
\includepdf[pages={1}]{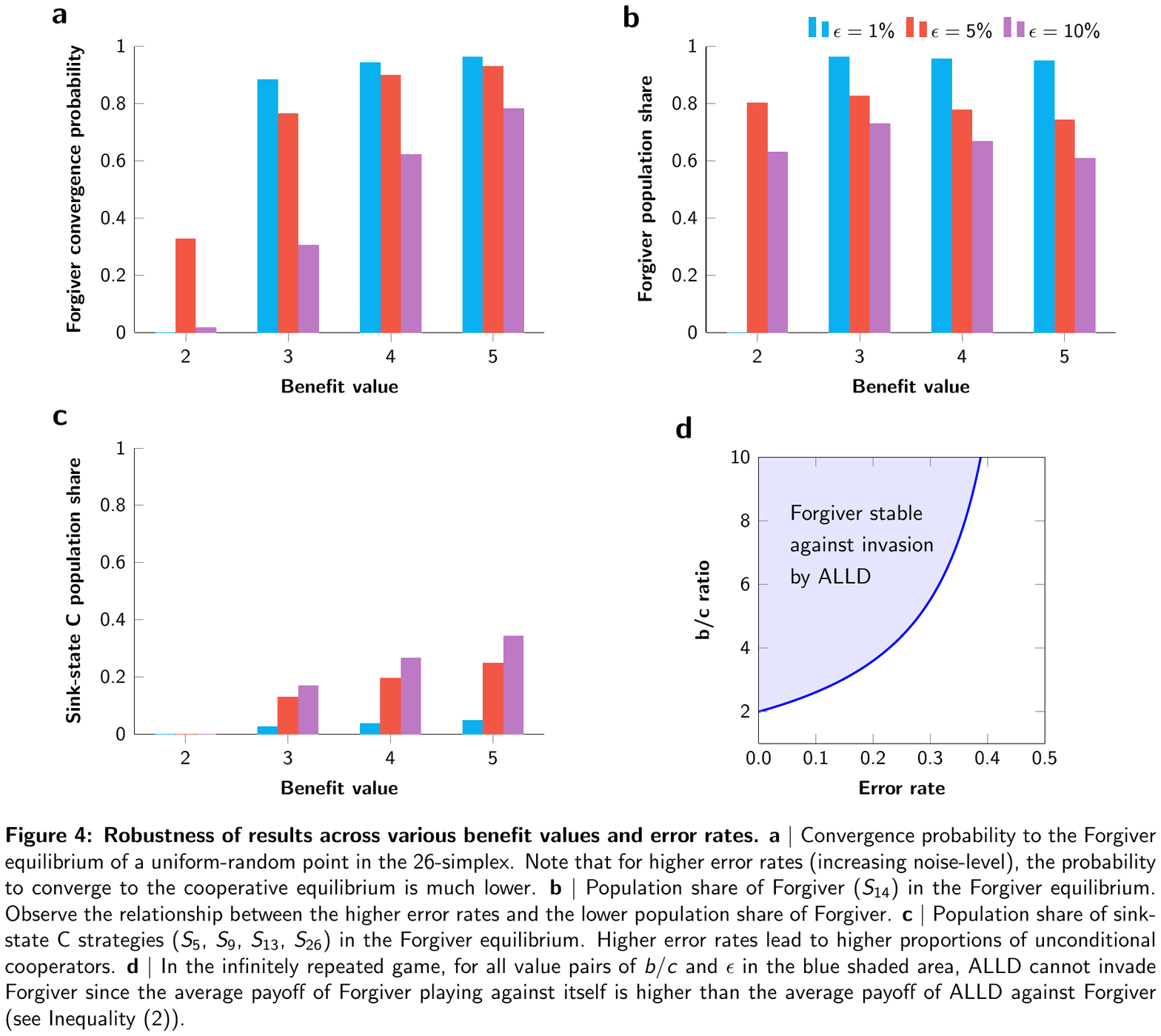}

\end{document}